\documentclass[12pt,a4paper]{article}
\usepackage{amsmath}
\usepackage{graphicx}
\usepackage{amsfonts}
\usepackage{amssymb}
\begin{document}

\begin{center}
\vspace{15mm}

{\textbf{{\Large {UNIVERSAL\ RELAXATION\ FUNCTION\ IN\ }}}}

\bigskip

{\textbf{{\Large {NONEXTENSIVE\ SYSTEMS}}} }

\bigskip

\textbf{F.Brouers and O.Sotolongo-Costa}\bigskip

Institute of Physics, Li\`{e}ge University and Henri Poincar\'{e} Chair of
Complex Systems, Havana University, Cuba.

fbrouers@ulg.ac.be , oscarso@ff.oc.uh.cu\bigskip

\textbf{Abstract}
\end{center}

\bigskip\baselineskip0.60cm

We have derived the dipolar relaxation function for a cluster model whose
volume distribution was obtained from the generalized maximum Tsallis
nonextensive entropy principle. The power law exponents of the relaxation
function are simply related to a global fractal parameter $\alpha$ and for
large time to the entropy nonextensivity parameter $q$. For intermediate times
the relaxation follows a stretched exponential behavior. The asymptotic power
law behaviors both in the time and the frequency domains coincide with those
of the Weron generalized dielectric function derived from an extension of the
Levy central limit theorem. They are in full agreement with the Jonscher
universality principle. Moreover our model gives a physical interpretation of
the mathematical parameters of the Weron stochastic theory and opens new paths
to understand the ubiquity of self-similarity and power laws in the relaxation
of large classes of materials in terms of their fractal and nonextensive properties.

\section{Introduction}

These last decades, growth, aggregation and fragmentation models have been
proposed and discussed with the aim of representing the (multi)-fractal
geometry and the resulting scaling properties of a great variety of complex
physical systems such as glasses, polymers, colloids, gels, self-similar
porous and cellular materials \cite{Jul88}. These studies cover a large range
of physical (electrical, dielectric, optical, magnetic and mechanical)
properties \cite{Ram87}. It has also been observed that universality is the
major feature of their scaling behavior.\cite{Jon77},\cite{Jon96}.

In the same context, it is more and more accepted that the Boltzmann-Gibbs
statistics is not adequate to describe the macroscopic thermodynamic
properties of natural phenomena when the effective microscopic interactions
and the microscopic memory are long ranged due to complex non-equilibrium
growth, aggregation or fragmentation processes. For that reason a generalized
form of a nonextensive entropy known as the ''Tsallis entropy'' \cite{Tsa88}
has been used with some success although the fundamental basis of such a
formulation is still the object of many discussions. Recently, Abe and
Rajagopal \cite{Abea00} have shown that Gibbs theorem for canonical ensemble
theory is not universal and that counting rule for statistical description can
depend on physical systems. This nonuniqueness can lead to various forms of
the entropy and to appropriate modifications of the maximum entropy principle.
An increasing number of papers are dealing with this problem \cite{Bar02}%
\cite{Lan02}.

Sotolongo et al. \cite{Sota00} starting from Tsallis'expression have used the
principle of maximum entropy with some natural constraints to derive a
fragment size distribution function in a regime in which long-range
correlations between all parts and scaling are present. This very general
principle gives rise to scaling without additional assumptions and the
resulting size distribution can be consider as a paradigm of physical systems
where multiscale interactions play the key role in determining the
hierarchical form of the fragment or cluster size distribution. It is the
reason why we have found interesting to use this distribution to analyze the
relaxation properties of complex materials.

Relaxation (dielectric, mechanical, magnetic...) in systems such as the ones
we are considering here is a complex stochastic mechanism which depends both
of the cluster geometric structure and the collective nature of the
interactions. The whole process of regression towards a steady
pseudo-equilibrium state when an external field is applied to or release from
the material is dynamical in nature. In most cluster models, relaxation is
viewed as a complex hierarchical or stochastic process\cite{Dis83}\cite{Wer97}.

We will view here a cluster as a set of units (atoms, molecules, aggregates)
relaxing collectively due to their interactions and their geometric structure.
The scale of the relaxation is time dependent.Small clusters relax first as a
result of intra-cluster interactions. For larger time due to long-range
interaction relaxing clusters are connected to the already relaxed clusters
and therefore depend on intra- and inte- cluster interactions.

We will consider the relaxation process from a macroscopic perspective and
introduced two global parameters one defining the time and space fractal
nature of the\ relaxation ($\alpha)$ and a second derived from the
maximization of the nonextensive entropy ($q)$ and characterizing the
hierarchical structure of the cluster geometry.In that way we aim at deriving
a macroscopic ''universal'' relaxation function valid for many physical
systems with different relaxation mechanisms at the micro and meso scale. We
consider here only electrical dipolar relaxation.

Indeed in that case, an universal pattern independent of materials and
microscopic model has been noticed for a long time \cite{Jon77},\cite{Jon96}
and widely discussed. Experimentally one observes both for large $t$ and small
$t$ a fractional power law dielectric response in the time domain :%
\begin{equation}
f(t)=-\frac{d\phi}{dt}=\left\{
\begin{array}
[c]{c}%
(\omega_{p}t)^{-n}\text{ \ for }\omega_{p}t<<1\\
\\
(\omega_{p}t)^{-m-1}\text{ for }\omega_{p}t>>1
\end{array}
\right.
\end{equation}
with $0<n,m<1.$ Quite generally it is observed that $1-n<m.$ The frequency
$\omega_{p}$ is material dependent. For intermediate $t$, the relaxation
function $\phi(t)$ is usually fitted to a stretched exponential, the so-called
Williams-Watts form\ : $\phi(t)\thicksim$ $\exp(-(\omega_{p}t)^{s}$. The small
$t$ (large $\omega$) exponent $n$ is related to small clusters relaxation and
the large $t$ (small frequency) exponent $m-1$ describes the relaxation of
large clusters. As it will appear in this work, they are not independent$.$
This non-Debye behavior has its counterpart in the frequency domain where
generalizations of the Cole and Cole dielectric functions with exponents
linked to $n$ and $m$ are commonly used by experimentalists.

The purpose of this letter is to show that our ($\alpha,q)$ model leads
naturally to this universal behavior and relates simply the exponents $n$ and
$m$ to the two parameters $\alpha$ and $q$. Asymptotically and numerically our
results coincide with the general dipolar relaxation function derived by Weron
and collaborators \cite{Wer97} using generalizations of the Levy central-limit
theorem. This unexpected result gives a physical meaning to the mathematical
parameters appearing in the purely stochastic Weron theory.

\section{ Cluster size distribution function}

Following references \cite{Sota00},\cite{Sotb00} we start from the expression
of the generalized entropy proposed by Tsallis \cite{Tsa88} and formulate the
maximization process as in \cite{Tsa01}%

\begin{equation}
S_{q}=k_{B}\frac{1-\int_{0}^{\infty}p^{q}(x)dx}{q-1}%
\end{equation}
The integral runs over all admissible values of the quantity $x$ and $x+dx$
and $p(x)dx$ is the probability of the system being in the state between $x$
and $x+dx$. The quantity $q$ is a real number characterizing the nonadditivity
of the entropy (nonextensivity) for system having nonexponential statistical
distribution and $k_{B}$ is the Bolzmann constant. When $q\rightarrow1,$ one
recovers the BG statistics $i.e$ $S_{q}\rightarrow S=-k_{B}\int_{0}^{\infty
}p(x)\ln p(x)dx.$

If we denote the volume of a cluster by $V$ and some typical volume $V_{m}$
characteristic of the distribution (for example the volume having the linear
dimension of a disorder or correlation characteristic length), we can define a
dimensionless volume $v=V/V_{m}$ which corresponds to the quantity $x$ in
eq.(2). The first constraint to impose is the normalization condition%

\begin{equation}
\ \int_{0}^{\infty}p(v)dv=1
\end{equation}

One has then to impose a mass conservation. As the system is finite and in
order to yield a slow decay in the asymptotic behavior of the cluster size
distribution, a more general condition, the ''q-conservation of the mass'' has
been imposed in the form%

\begin{equation}
\frac{\int_{0}^{\infty}vp^{q}(v)dv}{\int_{0}^{\infty}p^{q}(v)dv}=1
\end{equation}
and re-worked on the basis of the ''normalized q-expectation values''
introduced in \cite{Tsa98}\cite{Pra99}.

Eq.(3) and eq.(4) are the constraints to impose in order to derive the
$q$-dependent CSDF (cluster size distribution function) using the method of
Lagrange multipliers . The extremization calculation leads to the following
hierarchical CSDF defined for $1<q<2$:%

\begin{equation}
p(v)=[1-\frac{1-q}{2-q}v]^{\frac{1}{1-q}}%
\end{equation}
which has the same formal form as in \cite{Sota00},\cite{Sotb00} but with
different coefficients since here we have used the more correct normalized
$q$-expectation values using the so-called ''escort''
probabilities\cite{Tsa01}. This distribution has been used successfully to
account for the transition to scaling observed in the behavior of fragments in
the process of breaking \cite{Sota00} as well as more subtle effects as the
dimension cross-over between small and large fragment regions when thick clay
plates and glass are fractured \cite{Sotb00}. This is the cluster distribution
we will use to discuss dipolar relaxation in self-similar structured systems.

\section{Cluster relaxation function}

We will assume following a common practice that the observed macroscopic
non-exponential relaxation behavior can be interpreted in terms of a weighed
average of a Debye exponential relaxation decay $\exp[-t/\tau]$ with respect
to a distribution of relaxation time $p_{\tau}(\tau)$. In this formulation the
relaxation function that describes the polarization decay with time after a
steady polarizing electrical field has suddenly been removed, is written as:%
\begin{equation}
\phi(t)=\int_{0}^{\infty}p(\tau)\exp[-t/\tau]\mathrm{d}\tau
\end{equation}
\

To relate the cluster distribution eq.(5) to the relaxation rate distribution,
we have to make some assumptions on the variation of the relaxation time with
the number of relaxing units $N$ within a cluster.\ In a previous paper on
relaxation in percolation clusters \cite{Vaz97}, following a model of linear
polymers, we assumed, as a first approximation, that the cluster relaxation
time $\tau$ was proportional to the number of atoms (or the volume) in .the
percolating cluster and obtained in that case the proper long time scaling
behavior. For more complex systems due to intra- and inter-cluster
interactions and reasons discussed in the introduction, it is reasonable to
assume that the relaxing time increases faster than linearly with $N$. The
relation of $\tau$ with $N$ has been mainly discussed for the stress
relaxation in the phenomena of polymerization and sol-gel transition. For
example, in a simple Rouse model \cite{deG78} the cluster relaxation time has
been identified with the time scale for diffusion of the cluster over a
distance corresponding to its own size.\ Since in such model $\tau_{N}%
=R_{N}^{2}/D_{N}$ where$D_{N}\thicksim N^{-1}$ and $R_{N}$ is the radius of
gyration which scales with the fractal mass of the cluster as $R_{N}\thicksim
N^{1/d_{f}}$ , we get the relation $\tau=N^{\frac{d_{f}}{2+d_{f}}}.$ Although
each relaxation mechanism has its own specificity and is material dependent,
it is legitimate to assume quite generally that the relaxation time of a
volume made of $N$ relaxing elements scales as: \
\begin{equation}
\tau=v^{1/\alpha}%
\end{equation}
with $0<\alpha\leq1.$ The exponent $\alpha$ represents macroscopically the
''fractal'' geometric and dynamical nature of the relaxation dynamics in other
words the geometrical and dynamical ''exploration'' to use an expression
coined by de Gennes\cite{deG82} We will therefore assume a relaxation time
distribution of the form:
\begin{equation}
p_{\tau}(\tau)=[1-\frac{1-q}{2-q}\tau^{\alpha}]^{\frac{1}{1-q}}%
\end{equation}
To make the comparison with the Weron \cite{Wer97} stochastic theory easier we
will write the relaxation function in terms of the distribution of relaxation
rate $\beta=1/\tau$.\ We therefore write the relaxation function $\phi
_{\alpha,q}(t)$after the proper change of variable as a Laplace transform relation:%

\begin{equation}
\phi_{\alpha,q}(t)=\int_{0}^{\infty}p_{\beta}(\alpha,q)\exp[-t\beta
]\mathrm{d}\beta
\end{equation}
with%
\begin{equation}
p_{\beta}(\alpha,q)=\alpha(1/\beta)^{(1+\alpha)}[1-\frac{1-q}{2-q}%
(1/\beta)^{\alpha}]^{\frac{1}{1-q}}%
\end{equation}
This normalized function belongs to the basin of attraction of the Levy
distribution with a Levy tail exponent $\alpha.$ Weron $et$ $al.$\cite{Wer97}
have argued that the $\beta$ in eq.(10) are not individual rates but effective
relaxation rates corresponding to random relaxation representing the
macroscopic behavior of the material.

From the relaxation function one can then derive the relaxation response $\ $
$f(t)=-\frac{d\phi}{dt}$. The one--sided Fourier transform%

\begin{equation}
\chi(\omega)=\chi^{\prime}(\omega)-i\chi^{\prime\prime}(\omega)=\int
_{0}^{\infty}\exp(-i\omega t)f(t)dt
\end{equation}
yields the susceptibility response in the frequency range .If $\alpha=1,$ the
derivative of the Laplace transform eq.(9) gives:%

\begin{equation}
f_{\alpha=1,q}(t)=\Gamma\lbrack\frac{1}{q-1}]U[\frac{1}{q-1},1,\frac
{(1-q)t}{q-2}]
\end{equation}
$\Gamma$ is the Euler gamma function and $U(a,b,z)$ is the confluent
hypergeometric function : $\frac{1}{\Gamma(a)}\int_{0}^{\infty}e^{-zt}%
t^{a-1}(1+t)^{b-a-1}dt$. Using the asymptotic behavior of functions
$U(a,b,z),$ we get for $t\rightarrow\infty$ a power law.%

\begin{equation}
\lim_{t\rightarrow\infty}f_{\alpha=1,q}(t)\propto t^{-\frac{1}{q-1}}%
\end{equation}
when $t\rightarrow0,$ one gets a simple Debye behaviour.This first result
indicates that the high $t$ power law behavior is associated with the Tsallis
$q$ entropy parameter. When $0<\alpha\leq1,$ an analytical expression is not
available but an analysis of the results for some rational values of $\alpha$
and implying MeijerG functions allows us to conjecture the following
asymptotic behavior confirmed numerically for any couple of values ($\alpha
,q$) in the proper physical range.%

\begin{align}
\lim_{t\rightarrow0}f_{\alpha,q}(t)  &  \propto t^{\alpha-1}\\
\lim_{t\rightarrow\infty}f_{\alpha\text{ },q}(t)  &  \propto t^{-1-\alpha
\frac{2-q}{q-1}}%
\end{align}

We see that in this model, the empirical exponent $n$ and $m$ are simply
related to the two parameters $\alpha$ and $q$ : $n=1-\alpha$ and
$m=\alpha\frac{2-q}{q-1}.$ If $\alpha=1,$ one recovers the asymptotic behavior
of eq.13. For intermediate time, the response function can be fitted to the
derivative of a stretched exponential function with exponent $\alpha$ over
several orders of magnitude. The condition $0<m<1$ \ defines the range of
$q:\frac{1+2\alpha}{1+\alpha}<q<2.\;$The narrower range observed in most
experimental data $1-n<m<1$ defines a narrower physical range$:\frac
{1+2\alpha}{1+\alpha}<q<3/2.$

At this point it is of interest to compare our results with the only
self-contained mathematical theory which can account for the relaxation
universality in the framework of modern stochastic theory \cite{Wer97}. It has
been demonstrated mathematically \cite{Mon84},\cite{Wera93} that a stretched
exponential relaxation function can be obtained if and only if the relaxation
rate probability distribution is a completely asymmetric Levy distribution
function. In that case the low $t$ exponent can be related to the
Levy-$\alpha$ parameter i.e. $n=1-\alpha$, but this model does not yield the
large $t$ power law \cite{Wera93}. This results has been expressed as a
consequence of the Levy central limit theorem. By generalizing further the
Levy central limit theorem assuming a randomness in the number of effective
relaxation channels , Weron et al. \cite{Wer97} were able to derive using
precise stochastic theory arguments a general simple analytical form of the
relaxation function which reconcilliates and unifies the two power-laws and
the stretched exponential empirical expressions.

The Weron expression of relaxation function reads%

\begin{equation}
\phi_{\alpha}^{W}(t,k)=[1+k(\omega_{c}t)^{\alpha}]^{-1/k}%
\ \ \ \ \ \ \text{with\ }\ \ \ 0<\alpha\leq1,\ \ \ k\geq\alpha\
\end{equation}
The asymptotic behavior of the response function $f_{\alpha}^{W}%
(t,k)=-\frac{d\phi^{W}(t,k)}{\text{d}t}\ $for small and large $t$ is given by%

\begin{equation}
f_{\alpha}^{W}(t,k)=\left\{
\begin{array}
[c]{c}%
\alpha\omega_{c}(\omega_{c}t)^{\alpha-1}\text{ \ for }\omega_{c}t<<1\\
\\
\alpha\omega_{c}k^{-1-1/k}(\omega_{c}t)^{-\alpha/k-1}\text{ for }\omega
_{c}t>>1
\end{array}
\right.
\end{equation}
This corresponds with the results of our relaxation ($\alpha,q$) model if we
make the identification%

\begin{equation}
k=\frac{q-1}{2-q}%
\end{equation}
By Fourier transform it can then be shown straightforwardly from the two
asymptotic behavior in the time domain that both models obey the Jonscher
universal laws :%

\begin{equation}
\lim_{\omega\rightarrow\infty}\frac{\chi^{^{\prime\prime}}(\omega)}%
{\chi^{\prime}(\omega)}=\cot(n\pi/2)
\end{equation}

and%

\begin{equation}
\lim_{\omega\rightarrow0}\frac{\chi^{^{\prime\prime}}(\omega)}{\chi^{\prime
}(0)-\chi^{\prime}(\omega)}=\tan(m\frac{\pi}{2})
\end{equation}
with $n=1-\alpha,$ \ $m=\alpha/k,$ and $k=(\frac{q-1}{2-q}).$These relations
have a thermodynamic content \cite{Dis87} and the observation that $m$ is
related to a nonextensive entropy parameter has to be further analyzed.

\section{Numerical results and discussion}

In Fig.1, we show the behavior of the relaxation function $\phi_{\alpha,q}(t)$
for $\alpha=0.5$ and for $q$ varying in the range $1<q<2$. The slowing down of
the relaxation process increases with $q$.

Fig.2 shows the response function $f_{\alpha,q}(t)$ in a $\log$-$\log$ plot
for a typical value of $n=0.48$ ($\alpha=0.52)$ $,$ $m=0.68\ (q=1.433).$for
our model. and for the Weron response function $f_{\alpha,k}^{W}(t)$ with
$k=\frac{q-1}{q-2}$ in the same time units ($\omega_{p}=1)$. These values
correspond to a typical experimental example quoted and discussed in
\cite{Dis83} .The two models can fit the relaxation and dielectric behavior of
a large number of data reported in Jonsher \cite{Jon96} for dipole relaxation
systems. For values around $t=1$, $\ f(t)$ behaves as the time derivative of a
stretched exponential over 3 or 4 decades and again most of the experimental
data found in the literature with $\alpha$ ranging from to.$0.3.$ to $0.8$
\cite{Mon84} can be accounted for.

The main objective of this letter was to open a new line of thinking in this
problem and to show that general thermodynamic principles based on extension
of the concept of entropy appears to be a natural way to understand the
ubiquity of hierarchic self-similarity and power laws in the relaxation
behavior of large classes of materials.\ Moreover as demonstrated in
\cite{Abeb00} the power law behavior of statistical distribution is rooted in
the central limit theorem and a complete theory of relaxation based on
nonextensive arguments should be based on formal stochastic arguments. In this
letter we have shown that it is indeed the case since we found a simple
relation between the Tsalllis parameter $q$ and the stochastic parameter $k$
in the Weron generalized relaxation function. It should be emphasized that the
Tsallis generalization is not unique. Several alternative definitions of
entropy has been elaborated \cite{Bar02}\cite{Lan02}. The relation between $k$
and $q$ is therefore not unique \cite{Bar02}\cite{Lan02} and a general
discussion of the link of relaxation with nonextensivity will be presented in
further publications. We finally want to stress that the ideas developed in
this letter are not restricted to dipolar relaxation.

\section{Acknowledgments}

One of us (F.B.) is grateful to Prof.Karina Weron for illuminating
discussions.\ This work was supported by the ''Alma Mater'' project (Cuba) and
The Communaut\'{e} Wallonie-Bruxelles Action Concert\'{e}e 00/05-265\ (Belgium).

\begin{center}
\bigskip

{\large FIGURE CAPTIONS}
\end{center}

\bigskip

\noindent Fig1. Relaxation function $\phi_{\alpha,q}(t)$ for values $1<q<2.$
and $\alpha=0.5.$

\bigskip

\noindent Fig2. Response function $f_{\alpha,q}(t)=-\frac{d\phi_{\alpha,q}%
(t)}{dt}$ for $\alpha=0.52,q=1.433$ (full line)$.$ The dotted line is the
Weron $f_{\alpha,}^{W}(k,t)$ for $\alpha=0.52,k=0.765.\;$The exponents $\ n$
and $m$ are$.n=48$ \ and $m=0.68.$ We have used the relation $k=\frac
{q-1}{q-2}.$

\bigskip

\begin{thebibliography}{99}
\bibitem{Jul88}Julien R., Peliti L., Rammal R. and Boccara N.
\textit{Universalities in Condensed Matter } (Springer-Verlag) 1988.

\bibitem {Ram87}Ramakrishnan T.V. and Raj Lakshmi M., \textit{Non-Debye
Relaxation in Condensed Matter }(World Scientific Singapore) 1987.

\bibitem {Jon77}Jonscher A.K., \textit{Nature } \textbf{267} (1977) 673.

\bibitem {Jon96}Jonscher A.K., \textit{Universal Relaxation Law }%
(Chelsea-Dielectrics Press, London) 1996.

\bibitem {Tsa88}Tsallis C., \textit{J.of Stat.Phys.} \textbf{52}%
\textit{\ }(1988) 479

\bibitem {Abea00}Abe S. and Rajagopal A.K., \textit{Phys.Lett.} \textbf{A272}
(2000) 341.

\bibitem {Bar02}Baranger M.\textit{Physica} \textbf{A305} (2002) 27.

\bibitem {Lan02}Landsberg P.T.\textit{Physica} \textbf{A305} (2002) 32.

\bibitem {Sota00}Sotolongo-Costa O., Rodriguez Arezky H. and Rodgers G.J.,
\textit{Entropy} \textbf{2} (2000) 77.

\bibitem {Dis83}Dissado L.A. and Hill R.M. \textit{Proc.R.Soc.Lond}.\ A390
(1983), 134

\bibitem {Wer97}Weron K. and Kotulski M., \textit{J.of Stat.Phys.}
\textbf{88}\textit{\ }(1997)

\bibitem {Sotb00}Sotolongo-Costa O., Rodriguez Arezky H. and Rodgers G.J.,
\textit{Physica} \textbf{A286} (2000) 638.

\bibitem {Tsa01}Tsallis C. in \textit{Non extensive statistical mechanics and
thermodynamics} (Ed.S.Abe and Y.Okamoto, Springer Ed., Berlin 2001)

\bibitem {Tsa98}Tsallis C.,Mendes R.S.,and Plastino A.R. \textit{PhysicaA}
\textbf{A261 }(1998) 534.

\bibitem {Pra99}Prato D. and Tsallis C., \textit{Phys.Rev}. \textbf{E60}
(1999) 2398.

\bibitem {Vaz97}Vazquez A., Sotolongo-Costa O. and Brouers F.,
\textit{J.Phys.Soc.Japan} \textbf{66} (1997) 2324.

\bibitem {deG78}de Gennes P.G \textit{Comptes Rendus Acad. Sc. (Paris)} 226B
(1978) 131

\bibitem {deG82}de Gennes P.G. \textit{J.Chem.Phys}.76 (1982) 3316

\bibitem {Mon84}Montroll Elliott W and Bendler John T., \textit{J.of
Stat.Phys.} \textbf{34} (1984) 129

\bibitem {Wera93}Jurlewicz A. and Weron K., \textit{J.of Stat.Phys.}
\textbf{73} (1993) 69.

\bibitem {Dis87}Dissado L.A. and Hill R.M. Chemical Physics 111 (1987) 193.

\bibitem {Abeb00}Abe S. and Rajagopal A.K.\textit{Euro.Phyics Lett.
\textbf{52} (2000) 610}
\end{thebibliography}
\end{document}